\documentclass[conference]{IEEEtran}
\IEEEoverridecommandlockouts
\usepackage{cite}
\usepackage{amsmath,amssymb,amsfonts}
\usepackage{algorithmic}
\usepackage{graphicx}
\usepackage{textcomp}
\usepackage{xcolor}
\usepackage{subfigure}
\usepackage[misc,geometry]{ifsym}

\usepackage{times}  
\usepackage{helvet}  
\usepackage{courier}  
\usepackage{url}            
\usepackage{amsfonts}       
\usepackage{nicefrac}       
\usepackage{xcolor}         
\usepackage{mathrsfs}
\usepackage{microtype}
\usepackage{multirow}
\usepackage{booktabs}

\usepackage{amsmath, amssymb}
\usepackage{mathrsfs}
\usepackage{subfigure}
\usepackage{pgf-pie}
\usepackage{pgfplots}
\pgfplotsset{compat=1.18}
\usepackage{caption} 
\usepackage{algorithm}
\usepackage{algorithmic}
\def\BibTeX{{\rm B\kern-.05em{\sc i\kern-.025em b}\kern-.08em
    T\kern-.1667em\lower.7ex\hbox{E}\kern-.125emX}}
\begin{document}

\title{Learning Expressive Disentangled Speech Representations with Soft Speech Units and Adversarial Style Augmentation}

\author{\IEEEauthorblockN{Yimin Deng$^{1,2\ddagger}$, Jianzong Wang$^{1\ddagger}$\thanks{$^\ddagger$ Both authors have equal contributions.}, 
Xulong Zhang$^{1\textrm{\Letter}}$\thanks{$^\textrm{\Letter}$ 
Corresponding author: Xulong Zhang (zhangxulong@ieee.org).}, Ning Cheng$^{1}$, Jing Xiao$^{1}$}
\IEEEauthorblockA{$^{1}$Ping An Technology (Shenzhen) Co., Ltd., China\\$^{2}$University of Science and Technology of China}}

\maketitle

\begin{abstract}
Voice conversion is the task to transform voice characteristics of source speech while preserving content information. Nowadays, self-supervised representation learning models are increasingly utilized in content extraction. However, in these representations, a lot of hidden speaker information  leads to timbre leakage while the prosodic information of hidden units lacks use. To address these issues, we propose a novel framework for expressive voice conversion called ``SAVC'' based on soft speech units from HuBert-soft. Taking soft speech units as input, we design an attribute encoder to extract content and prosody features respectively. Specifically, we first introduce statistic perturbation imposed by adversarial style augmentation to eliminate speaker information. Then the prosody is implicitly modeled on soft speech units with knowledge distillation. Experiment results show that the intelligibility and naturalness of converted speech outperform previous work.
\end{abstract}

\begin{IEEEkeywords}
voice conversion, speech synthesis, style augmentation, prosody modeling
\end{IEEEkeywords}

\section{Introduction}
\label{sec:intro}
Voice conversion~(VC) targets at transforming speaker style without changing content. It covers a wide range of applications in real life such as intelligent security products, video dubbing, and customer service.

A common strategy for handling VC tasks involves decoupling content information from speaker-related attributes in speech.
and manipulate these speech attributes. AutoVC~\cite{qian2019autovc} proposes the first autoencoder framework of zero-shot VC which encourages encoders to disentangle the content and timbre. Derived from AutoVC, Vector Quantization~(VQ)~\cite{VQVC,wu2020vqvc+} and contrastive learning are applied to VC models~\cite{tang2022avqvc,cln-vc} and improve conversion quality.

Recently, Self-Supervised Learning (SSL)~\cite{vq-wav2vec,chen2022wavlm,hsu2021hubert} has revolutionized research in AI areas. With complex layers structure and huge number of parameters, SSL models are able to capture rich contextual information contained in SSL feature. Through complex knowledge extraction in the pre-training stage on large-scale datasets, SSL feature can also be well applied to a variety of downstream tasks in speech domain~\cite{mohamed2022self,wang2021self,tsai2022super}. In the VC area, SSL models contribute a lot to content modeling~\cite{baevski2020wav2vec,chen2022wavlm,hsu2021hubert}. Taking a speech as input, SSL will generate speech units which can well represent linguistic information.

In the present day, the significance of converted speech extends beyond its naturalness to encompass its expressiveness as well.
Prosody modelling~\cite{ning2023expressiveVC,deng2023PMVC} is necessary for expressive voice conversion which derives downstream tasks like emotion computation and intent prediction.
Prosody includes pitch, energy and rhythm which express specific situation and emotion of speakers. Similar to linguistic information, prosodic information can be considered as time variant style and also should be independent with speakers. Hence, to model a speaker style more accurately, timbre and prosody should be both taken into account. 
Improvements to expressive VC are heading towards better timbre conversion and prosody modeling ability.

However, an imbalance exists in the utilization of SSL feature during voice conversion. Much residual speaker information in these units degrades timbre similarity in conversion while the prosodic information lacks use. The mainstream solutions for timbre leakage are based on information bottleneck~\cite{speechflow,lee2021voicemixer} and perturbation~\cite{qian2022contentvec,ning2023expressiveVC} respectively. Bottleneck structure reduces the quality of the content to some extent. Previous perturbation-based work relies on signal processing techniques like formant shifting to destroy timbre information and use the perturbed features as input~\cite{ning2023expressiveVC,choi2022nansy++}. However, the shifting scale factors need manual efforts during training. Recent studies~\cite{li2022uncertainty,zhong2022adversarial,zhang2023advstyle} found that uncertainty modeling with statistic perturbation helps to improve the model generalization in image style transfer, possible to be a solution for timbre leakage. 

To address these issues, a novel framework named ``SAVC" is proposed which is based on Soft speech units and Adversarial style augmentation for Voice Conversion. 
Given soft speech units from HuBert-Soft~\cite{van2022softvc} as input, we design an attribute encoder to extract the time-variant information including content and prosody respectively. In particular, we impose statistic perturbation on speech style with adversarial augmentation to eliminate speaker information. It encourages the attribute encoder to learn similar time-variant features under such perturbation of speaker style. Then, to learn the disentangled prosody feature, knowledge distillation is introduced. The main contributions of this paper can be formulated as follows:

\begin{enumerate}
    \item We use soft speech units from HuBert and apply them into both content and prosody modeling for expressive voice conversion.
    \item We propose an adversarial style augmentation to impose dynamic statistic perturbation on features to reduce timbre leakage on hidden units.
    \item We apply knowledge distillation in prosody modeling. The student model can learn expressive prosody embedding without requiring the explicit prosodic features.
\end{enumerate}

\section{Related Work}
\subsection{Voice Conversion}
Voice conversion is one of significant downstream tasks for disentangled speech representation learning~\cite{qian2022contentvec}. Better disentanglement of different voice characteristic leads to better sound quality of converted speech. In previous work~\cite{tang2021TGAVC}, taking traditional acoustic features like mel-spectrum as input, the cnn-based content encoders need to design bottleneck carefully to discard speaker-related information~\cite{qian2019autovc,VQVC,park2023triaan}. AutoVC~\cite{qian2019autovc} is the first to propose auto-encoder framework for zero-shot voice conversion. To ensure the VC performance, the derived work needs to follow similar bottleneck structure and disentangle content information. Vector Quantization~(VQ) is introduced to transfer continuous feature into discrete feature~\cite{VQVC,wu2020vqvc+} as another kind of bottleneck and discards speaker-related information to some content. 
    
Recently, the self-supervised learning models have aroused more and more public attention. 
These SSL models exhibit strong performance in downstream tasks, including but not limited to speech recognition, speaker verification, and voice conversion~\cite{mohamed2022self,wang2021self,tsai2022super}. They are pretrained with large scale of corpus and able to capture rich contextual information from waveform. FreeVC~\cite{li2023freevc} uses WavLM~\cite{chen2022wavlm} to extract content information through a bottleneck extractor. ContentVec~\cite{qian2022contentvec} adopts a similar structure of the HuBert and distills a model to extract content information for voice conversion. 
As a compromise between continuous raw features and discrete labels preserves more content and less speaker information, Soft-VC~\cite{van2022softvc} drops out the quantization of HuBert and uses soft speech units from a linear layer to improve the intelligibility. 

\subsection{Timbre Leakage}
The natural transition of speaker identity signifies the success of voice conversion tasks. In order to capture the rich perceptual content of speech from shallow to deep levels, some SSL models choose a multi-layer transformer structure. It was found that the lower layers of SSL models tend to learn speaker-related information while the upper layers are tasked with encoding content-related information~\cite{wang2022inter}. It's unavoidable for speech units to contain much timbre information from source speakers. Timbre leakage from source speaker degrades the voice similarity to the target, leading to the failure of voice conversion. The mainstream solutions can be divided into two categories: information bottleneck-based and perturbation-based method. To focus more on content modeling, HuBert~\cite{hsu2021hubert} consider quantization layer as bottleneck to discard speaker information. On the other side, ContentVec~\cite{qian2022contentvec} adopts the random perturbation on audio features~\cite{choi2021nansy} which improve the sound quality and naturalness. Zhao et.al~\cite{advhubert} removes speaker-related detail adversarially by using a special discriminator. 

It has been observed that feature statistics characterize the style information in various style transfer~\cite{huang2017arbitrary,chou2019invc}. Inspired by the statistics perturbation~\cite{nuriel2021permuted}, Zhong~\cite{zhong2022adversarial} et.al and Zhang~\cite{zhang2023advstyle} et.al apply Adversarial Style Augmentation to transfer images style without extra discriminators. It's observed that such statistics perturbation can generate style augmentation samples and has potential power for voice style transfer.

\subsection{Prosody Modeling}
Prosody modeling have become important considerations for the performance of the VC models. Integrating prosody modeling becomes essential to learn prosody representation and capture the nuances of intonation and prosody~\cite{weston2021learning,deng2023PMVC}. Recent studies~\cite{lin2023utility} found that the latent prosodic information from SSL models shows good performance in prosody pseudo tasks and prosody-intensive tasks. With additional sequence-modeling layers, SSL models can be able to predict some perceptions of speaking style. However, few work take advantage of the potential prosody modeling power of SSL models. That leads to a waste of such rich prosodic source. 

To address the issues, we propose a novel framework named ``SAVC" based on Soft speech units with Adversarial style augmentation. Specially, an attribute encoder is forced to generate similar speaker-independent features for style augmented samples. Besides, to disentangle the prosodic feature with content feature, a teacher model with pre-trained prosody encoder is employed.


\section{Methodology}
\subsection{Model Structure}
Inspired by the various applications of SSL models in speech domains, we explore the utilization of soft speech units in VC. HuBert-Soft takes the raw waveform as input and produces soft speech units. Compared to discrete units produced by HuBert-Base~\cite{hsu2021hubert}, soft units capture more speaker-independent information and bring notable improvement in intelligibility and naturalness. 

\begin{figure*}[!t]
    \centering
    \includegraphics[scale=0.5]{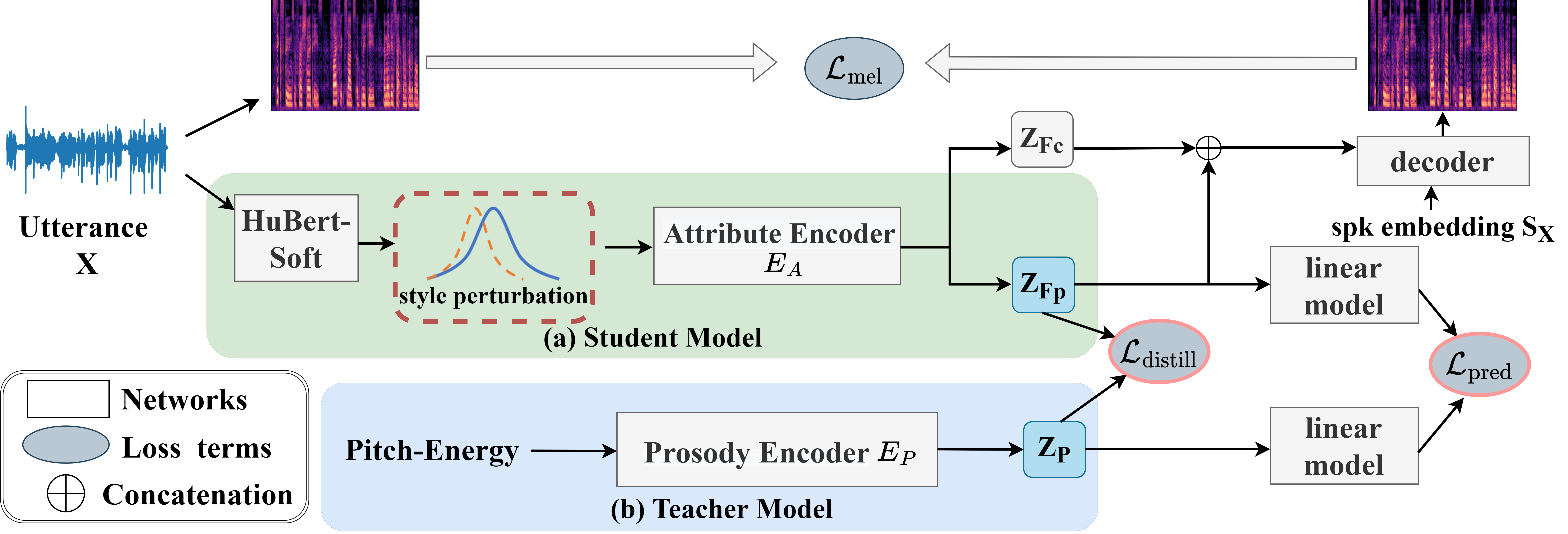}
		 
	
	%
    \caption{The framework of SAVC. $Z_{F_c}$, $Z_{F_P}$ are the content embedding and prosody embedding extracted by the attribute encoder respectively. $Z_P$ is the prosody embedding extracted by the teacher model. Speaker embedding $S_X$ is extracted by a pre-trained model.}
    \vspace{-1.5em}
    \label{fig:pip}
\end{figure*}

The framework of the SAVC model includes the following modules as illustrated in Fig.~\ref{fig:pip}. An attribute encoder $E_A$, extracting content embedding $Z_{Fc}$ and prosody embedding $Z_{Fp}$ from input soft speech units. We impose speaker style perturbation on this input by designing an Adversarial Style Augmentation illustrated in Fig.~\ref{fig:asa}. Besides, the d-vector from target speaker extracted by a pre-trained model~\cite{d-vector} is used to provide the speaker embedding $S_X$ as timbre reference. The explicit prosody encoder $E_P$ in the teacher model maps prosodic features into latent space and we train the student model with feature distillation losses. 
The decoder is trained to reconstruct the input speech utilizing extracted content embedding, prosody embedding, and speaker embedding.

\subsection{Speaker Irrelevant Attributes Extraction}
The leakage of residual speaker information in soft speech units still remains challenging which may degrade VC performance in voice similarity to target. The perturbation-based solution expects that the attribute encoder can learn similar representations even though the inputs vary in different style perturbation.
Instead of signal processing like formant shifting~\cite{choi2022nansy++}, we propose a novel dynamic statistic perturbation to destroy the source speaker style.

\subsubsection{Feature statistic modeling}
Following the description in INVC~\cite{chou2019invc}, we can perform style transformation on features by statistic replacement as:

\begin{equation}
    s_t = \sigma_t(\frac{s-\mu(s)}{\sigma(s)})+\mu_t
\end{equation}
where $\mu(s)$ and $\sigma(s)$ denotes the statistics of source features $s$, $s_t$ are features for new styles decided by the factors $\mu_t$ and $\sigma_t$. 
Traditional approaches commonly incorporate $\mu_t$ and $\sigma_t$ using feature statistics within a mini-batch, which involves randomly selecting a segment from the utterance. However, such methods constrain the representational capacity of $\mu_t$ and $\sigma_t$ within a limited statistical space. Rather than treating each feature statistic $\mu_t$ and $\sigma_t$ as deterministic values measured from the learned feature, uncertainty modeling~\cite{li2022uncertainty} considers that they follow a multi-variate Gaussian distribution and they can describe the uncertainty scope for different potential style shift. It's assumed that $\mu_t \sim N(\mu(s),\Sigma_\mu)$ and $\sigma_t \sim N(\sigma(s),\Sigma_\sigma)$ in a batch.

\begin{figure}[htbp]
    \vspace{-1em}
    \includegraphics[scale=0.62]{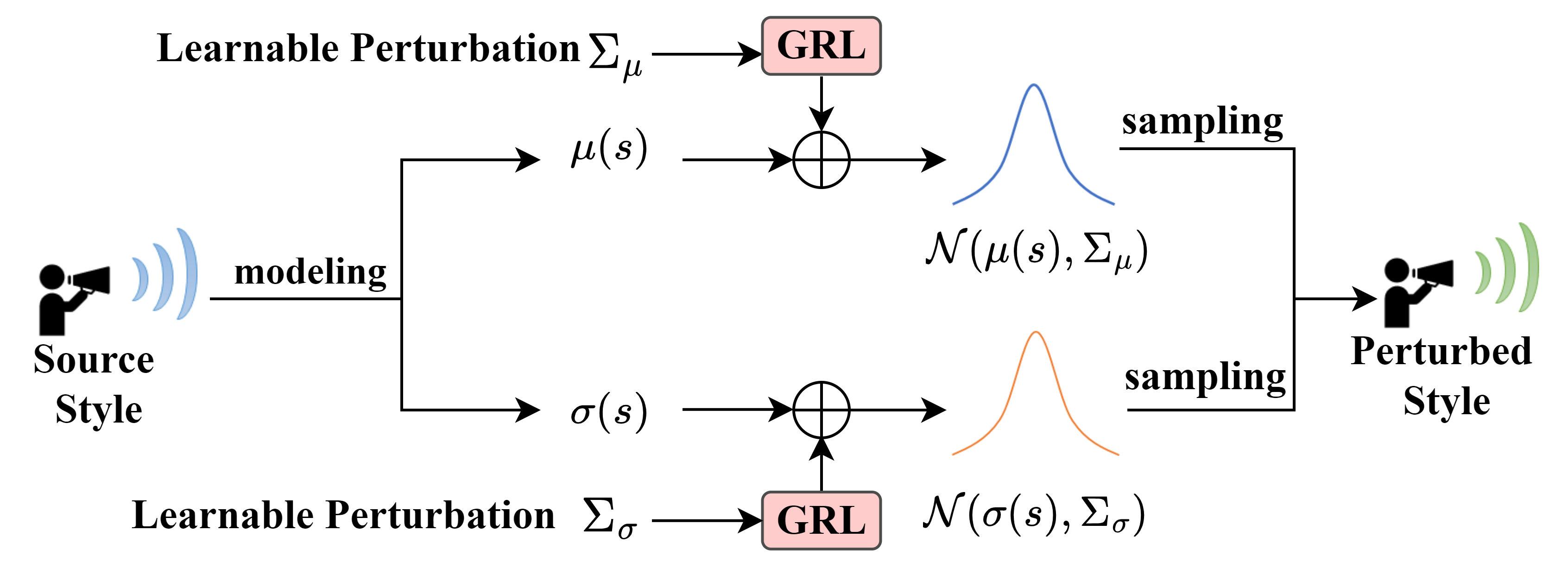}
	
	%
    \caption{Feature statistic based Adversarial Style Augmentation (ASA) module.}
    \label{fig:asa}
\end{figure}
\subsubsection{Adversarial Style Augmentation module}
First given soft speech units as input, before fed into the encoder, we design an Adversarial Style Augmentation~(ASA) module to perturb source speaker style as shown in Fig.~\ref{fig:asa}. Considering soft units as intermediate features learned from raw audio, we denote $s \in \mathbb{R}^{B \times W \times C}$~(B for batch size, W for the number of units, C for feature dimensions) as soft units. And denote $\mu \in \mathbb{R}^{B \times C}$ and $\sigma \in \mathbb{R}^{B \times C}$ as the mean and standard deviation of features in each channel of instance. As described in INVC~\cite{chou2019invc}, speaker style can be modeled with $\mu$ and $\sigma$. In the ASA module, we design dynamic statistic perturbation to destroy the source speaker style.

As shown in \textbf{Algorithm}~\ref{alg:algorithm}, given a batch of soft speech units as input, we extract the feature statistics of each instance and conduct normalization. Then construct the standard deviations $\Sigma_\mu$ and $\Sigma_\sigma$ of feature statistics $\mu(s)$ and $\sigma(s)$ for each instance. Use them to initialize learnable perturbation parameters. Then perform style transformation on each normalized instance with new feature statistics $\mu_t$ and $\sigma_t$. Considering the learnable perturbation parameters have opposite optimization target against the attribute encoder, a Gradient Reversal Layer~(GRL) is positioned between step 7 and step 8 to 
reverse the gradients in the back-propagation. Hence, the learnable 
style parameters become more influential in style representation space while the attribute encoder generates the similar output as possible under dynamic style perturbation in the training process.
\begin{algorithm}[tb]
\caption{Dynamic Style Perturbation Algorithm}
\label{alg:algorithm}
\begin{flushleft}
\textbf{Input}: A batch $\boldsymbol{S}$ of soft speech units $\boldsymbol{s}$\\
\textbf{Output}: A batch $\boldsymbol{S^{\prime}}$ of perturbed instances 
$\boldsymbol{s^{\prime}}$\\
\end{flushleft}
\begin{algorithmic}[1] 
\FOR {each instance $s \in \boldsymbol{S}$}
    \STATE Compute the mean $\mu(s)$ and standard deviation $\sigma(s)$ in each channel for an instance $s$ 
    \STATE Add normalized instance $s_n$ to $\boldsymbol{S_n}$
\ENDFOR
\STATE Construct $\Sigma_\mu$ and $\Sigma_\sigma$ as the standard deviations of $\mu(s)$ and $\sigma(s)$ of current batch. 
\STATE Initial learnable parameters $\mathcal{I}_\mu \leftarrow \Sigma_\mu $ and $\mathcal{I}_\sigma 
\leftarrow \Sigma_\sigma$
\STATE Construct the perturbed standard deviations $\Sigma_\mu^{\prime} = \frac{\mathcal{I}_\mu}{\mathcal{I}_\mu * \Sigma_\mu }$ , $\Sigma_\sigma^{\prime} = \frac{\mathcal{I}_\sigma}{\mathcal{I}_\sigma * \Sigma_\sigma }$
\FOR {each normalized instance $s_n \in \boldsymbol{S}_n$}
    \STATE Sample $\mu_t \sim \mathcal{N}(\mu(s),\Sigma_\mu^{\prime})$
    \STATE Sample $\sigma_t \sim \mathcal{N}(\sigma(s),\Sigma_\sigma^{\prime})$
    \STATE Perform style transformation $s^{\prime} = s_n * \sigma_t + \mu_t$
    \STATE Add $s^{\prime}$ to $\boldsymbol{S}^{\prime}$ \\
\ENDFOR

\end{algorithmic}
\end{algorithm}

\subsection{Prosody Modeling}
Prosody modeling relates to the expressiveness of generated speech from conversion model. In previous work that do not consider the disentanglement of prosodic represenation, the timbre can be changed but the prosody of generated speech still be same as the source speaker. In the proposed method, with extracted speaker irrelevant features, we can make full use of rich prosodic feature from soft speech units.
\subsubsection{Prosody Distillation}
Though the soft speech units is at a level between continuous state and discrete state, some continuous information may be discarded. Then unnatural sudden change in tone may occur. Typically, models do not inherently learn to generate smooth solution. It's challenging for soft units-based method to model prosody. Hence, we need an accurate teacher model to force the student to learn natural prosody implicitly.  

To avoid the fact that unmatched prosodic information may lead to misunderstanding of linguistic content and degrade the sound quality~\cite{speechflow}, we train the prosody modeling ability using knowledge distillation. Explicit prosody modeling refers to training an encoder to learn prosodic information from prosody features. Specially, for explicit and accurate modeling, we use two types of prosody features: pitch contour and energy. 
\begin{itemize}
    \item \textbf{Pitch:} We use the log scale of pitch value extracted by a toolkit \textit{pyworld}.
    \item \textbf{Energy:} We use a universal audio process toolkit \textit{librosa} to compute this value, and also use the log scale. 
\end{itemize}

As shown in Fig.~\ref{fig:pip}, the teacher model contains a pre-trained prosody encoder that receives normalized pitch and energy value and learn the prosody embedding. 
The attribute encoder learns prosody implicitly by the guidance of teacher model without explicit prosody feature input. Inspired by PMVC~\cite{deng2023PMVC}, we use one attribute encoder to extract both content and prosody features based on flexible bottlenecks. Considering there is no linguistic input in teacher model, the prosody feature extracted from student model can learn as less content as possible. During training, the teacher model is frozen and the student model is trained with a feature distillation loss between $Z_{F_P}$ and $Z_P$: 
\begin{equation}
    \mathcal{L}_{dis} = MSE(Z_{F_P},Z_P)
\end{equation}

\subsubsection{Expressiveness Constraints}
Moreover, it's observed that to evaluate the performance for SSL models in prosody-related task, additional sequence model layers are necessary~\cite{lin2023utility}. Inspired by this, to improve the expressive performance, we add a linear model with a downstream task at fine-tune stage. The prosody-related tasks include prosody-pseudo tasks and prosody-intensive tasks. During our experiment, we found that the model is fine-tuned with prosody-related tasks such as prosody reconstruction and prosody context prediction will result in obvious deviation leading to oscillatory training. So we focus on another general semantic prosody task. 

In available open-source emotional dataset, the utterances share the same content but with different emotions. Different emotions contain various prosody style. 
Specially, we use the prosody features from both teacher model and student model to predict emotion labels with one-hot encoding. The loss is marked as $\mathcal{L}_{pred}$ and the L2 loss is employed:
\begin{equation}
    \mathcal{L}_{pred} =\left\|y-\hat{y_t}\right\|_2^2+\left\|y-\hat{y_s}\right\|_2^2
\end{equation}
where $\hat{y_*}$ is the predicted vectors from teacher or student, and $y$ is the real label. With teacher guidance and fine-tuning strategy, the expressiveness of extracted prosody is improved and less meaningless information is generated from the attribute encoder.

\begin{table*}[!t]
  \caption{Subjective results in Many-to-Many VC and Zero-Shot VC}
  \centering
  \fontsize{8.7}{7}\selectfont
  \label{Comparison}
    \begin{tabular}{ccccccc}
    \toprule
    \multirow{2}{*}{\textbf{Methods}}&
    \multicolumn{3}{c}{\textbf{Many-to-Many VC}}&\multicolumn{3}{c}{\textbf{ Zero-Shot VC}}\cr
    \cmidrule(lr){2-4} \cmidrule(lr){5-7}
    & MOS$\uparrow$ & TSS$\uparrow$ &PSS$\uparrow$ & MOS$\uparrow$ & TSS$\uparrow$ &PSS$\uparrow$\cr
    \midrule
    INVC~\cite{chou2019invc} & 3.02 $\pm$ 0.33
    & 3.29 $\pm$ 0.17 &2.62 $\pm$ 0.12 
    & 2.87 $\pm$ 0.24 & 3.19 $\pm$ 0.21 &2.52 $\pm$ 0.33\cr
    UUVC~\cite{chen2023uuvc} & 3.56 $\pm$ 0.34
    & 3.86 $\pm$ 0.16 & 3.64 $\pm$ 0.08  
    & 3.27 $\pm$ 0.15 & 3.42 $\pm$ 0.20  & 3.48 $\pm$ 0.13\cr
    SpeechSplit2.0~\cite{chan2022speechsplit2} 
    & 2.86 $\pm$ 0.24
    & 3.03 $\pm$ 0.08 & 3.28 $\pm$ 0.14 
   & 2.62 $\pm$ 0.27 & 2.71 $\pm$ 0.31 &3.15 $\pm$ 0.28 \cr
    PMVC~\cite{deng2023PMVC} & 3.52 $\pm$ 0.14
    & 3.66 $\pm$ 0.28 & 3.58 $\pm$ 0.16 & 3.38 $\pm$ 0.10 & 3.55 $\pm$ 0.22 & 3.46 $\pm$ 0.24 \cr
    \midrule
    \textbf{SAVC} &\textbf{3.89 $\pm$ 0.22}
    &\textbf{3.93 $\pm$ 0.15} &\textbf{3.75 $\pm$ 0.17} 
    &\textbf{3.75 $\pm$ 0.10} &\textbf{3.87 $\pm$ 0.16} &\textbf{3.50 $\pm$ 0.22} \cr
    \bottomrule
    \end{tabular}
\end{table*}

\subsection{Training Strategy}

With extracted content feature $Z_{F_C}$, prosody feature $Z_{F_P}$ and speaker embedding $S_X$ from pre-trained d-vector model, the decoder generates reconstructed mel-spectrum of speech $X'$ during training. A reconstruct loss between the mel-spectrums $M( \cdot )$ of source speech $X$ and reconstructed speech $X'$ is computed as:
\begin{equation}
    \mathcal{L}_{rec} = \sum^{N}||M(X')-M(X)||_2^2
\end{equation}
where N denotes the number of speech during training.

It's expected that a pre-trained speaker encoder provides the target timbre representation and the other parts of ``SAVC" should be speaker-independent.
The Adversarial Style Augmentation can generate feature statistic-perturbed samples sharing the same content and prosody. Then the attribute encoder is encouraged to learn similar speaker-independent information from these samples.

Moreover, to disentangle prosody from content, we introduce the knowledge distillation. During pre-training process, the teacher model is independent with linguistic information. With the guidance of pre-trained prosody encoder, the latent linguistic information in prosody embedding $Z_{F_p}$ from attribute encoder will be reduced. 

During training, the objective function seeks to minimize a combination of the aforementioned terms, each weighted by hyper-parameters:
\begin{equation}\label{eq:total}
    \mathcal{L}_{total} = \alpha \mathcal{L}_{rec} + \beta\mathcal{L}_{dis} + \lambda\mathcal{L}_{pred}
\end{equation}

\subsection{Architecture of Attribute Encoder}
\begin{figure}
    \centering
    \includegraphics[scale=0.8]{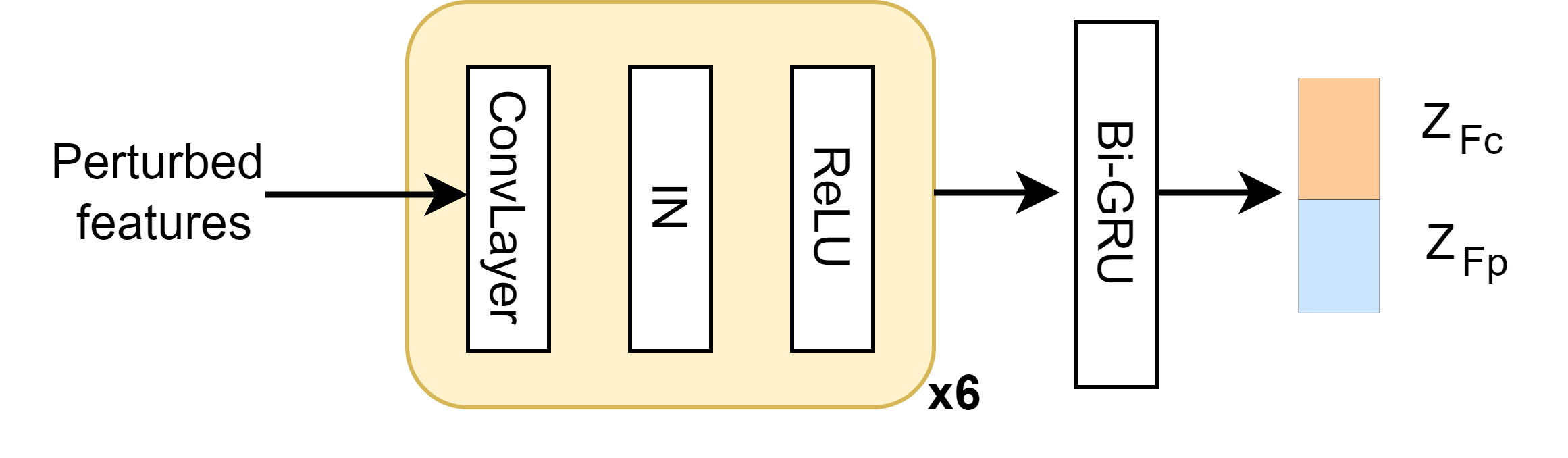}
    \caption{Architecture of attribute encoder}
    \label{fig:arc}
\end{figure}

In the proposed model, the attribute encoder is designed as shown in Fig.~\ref{fig:arc} with flexible bottleneck sizes like PMVC~\cite{deng2023PMVC}. Taking the perturbed features as input, the attribute encoder uses dilated convolution blocks and RNN to estimate the speaker-independent embeddings. Specially, we utilize Bi-GRU to reduce the number of parameters. In the teacher model, the pre-trained prosody encoder adopt a similar structure with extra attention module. It takes the final GRU state as query and a group of trainable style tokens as key and value. Opposite to the student model, the teacher model will take the explicit prosody features (pitch and energy values) as input.

\section{Experiment}

\subsection{Experiment Setup}
We train our model first on multi-speaker English corpus VCTK~\cite{vctk}, about 30 hours of recordings recorder by 99 speakers. Then fine-tune the prosody module on recordings of English speakers from another multi-speaker dataset with five kinds of emotion named Emotional Speaker Dataset~(ESD)~\cite{zhou2021esd}. We select the recordings of 6 English speakers (3 female and 3 male) as training dataset from ESD, and left four English Speakers for zero-shot VC tests. 

All recordings are resampled to 16 kHz. To generate 80-channel mel-spectrums, use a short-time Fourier transform (STFT) with Hanning window. Choose 1024 as FFT size, 1024 as window size and 256 as hop size. 
Set the weights in Eq.(\ref{eq:total})to $ \alpha = 1, \beta = 0.5, \lambda = 0.1$. 

To thoroughly assess the performance of SAVC, we undertake comparative experiments on various VC tasks. Many-to-many VC involves source and target speakers in the training dataset, while Zero-Shot VC involves unseen speakers during training.
We select the following models as baseline out of the following consideration:
\begin{itemize}
    \item \textbf{INVC~\cite{chou2019invc}:} It characterize style with determined feature statistic and adopts instance normalization to eliminate speaker information. 
    \item \textbf{UUVC~\cite{chen2023uuvc}:} It uses discrete HuBert units and explicit prosody modeling for expressive voice conversion. The prosody is modeled based on the input of pitch and energy.
    \item \textbf{SpeechSplit2.0~\cite{chan2022speechsplit2}:} It utilizes triple information bottlenecks with effective signal processing techniques. Detaily, we select the pre-trained model with wider bottlenecks for best performance.
    \item \textbf{PMVC~\cite{deng2023PMVC}:} PMVC uses one encoder to learn both content and prosody features. Based on random perturbation on prosody, the expressive voice conversion is realized with implicit prosody modeling.
\end{itemize}
We follow the training description in ~\cite{chou2019invc,chen2023uuvc,chan2022speechsplit2,deng2023PMVC}. To be fair, we use a pre-trained vocoder named Hifi-gan~\cite{kong2020hifi} to transform all the output mel-spectrogram into the waveform. 

\begin{table*}[!t]
  \caption{Objective results in Many-to-Many VC and Zero-Shot VC}
  \centering
  \fontsize{8.7}{7}\selectfont
  \label{objComparison}
    \begin{tabular}{ccccccc}
    \toprule
    \multirow{2}{*}{\textbf{Methods}}&
    \multicolumn{3}{c}{\textbf{Many-to-Many VC}}&\multicolumn{3}{c}{\textbf{ Zero-Shot VC}}\cr
    \cmidrule(lr){2-4} \cmidrule(lr){5-7}
    & MCD$\downarrow$ & CER$\downarrow$ & SES$\uparrow$  & MCD$\downarrow$ & CER$\downarrow$ & SES$\uparrow$ \cr
    \midrule
    INVC~\cite{chou2019invc} & 9.42 $\pm$ 0.11 & 35.5\% & 75.5\%
    & 9.63 $\pm$ 0.23 &36.8\% &67.1\% \cr
    UUVC~\cite{chen2023uuvc} & 5.82 $\pm$ 0.12 & 12.2\% & 82.6\%
    & 6.20 $\pm$ 0.17 & 16.1\% & 79.8\% \cr
    SpeechSplit2.0~\cite{chan2022speechsplit2} & 7.80 $\pm$ 0.23 
    & 48.2\% & 73.3\%
    & 9.21 $\pm$ 0.13 & 52.5\% & 69.0\% \cr
    PMVC~\cite{deng2023PMVC}& 6.31 $\pm$ 0.09 & 15.5\% & 81.7\%
    & 6.64 $\pm$ 0.12 & 18.2\% & 80.6\% \cr
    \midrule
    GT &- &7.1\% & 98.6\%
    &- &7.6\% & 98.6\% \cr
    \textbf{SAVC} &\textbf{5.19 $\pm$ 0.15} &\textbf{10.2\%} & \textbf{86.2\%}
    &\textbf{5.82 $\pm$ 0.20} &\textbf{11.7\%} & \textbf{81.0\%} \cr
    \bottomrule
    \end{tabular}
\end{table*}
\subsection{Subjective Evaluation}

12 volunteers, consisting of 6 males and 6 females, are invited to participate in subjective tests. The Mean Opinion Score (MOS) test is utilized to assess the naturalness of  converted speech from different models. Following previous work~\cite{deng2023PMVC}, the Voice Similarity Score (VSS) test is divided into Timbre Similarity Score (TSS) and Prosody Similarity Score (PSS).  Groups of utterances are rated for voice similarity on a scale ranging from 1 to 5.
They need to concern timbre and prosody information in TSS and PSS tests respectively. Both MOS and VSS are higher for better.  

As seen in Tab.~\ref{Comparison}, the proposed model achieves a lower TSS and PSS scores for less distortion than baselines. Both the timbre and prosody similarity is improved with a substantial degree of naturalness. Even in zero-shot condition, the proposed model has better robustness because of less performance decay compared to baselines.  


\subsection{Objective Evaluation}


Choose Mel-Cepstral Distortion (MCD) as an objective metric to quantify the disparity between the acoustic features of converted speech and the target speech which is lower to be the better. To measure the intelligibility of transformed speech, a popular ASR model named FunASR~\cite{gao2023funasr} is employed to measure Character Error Rate~(CER) as shown in Tab.~\ref{objComparison}. 
Besides, Speaker Verification (SV) assesses whether the converted utterance aligns with the speaker who provided the timbre information in the Voice Conversion (VC) process.
So we utilize a toolkit WeSpeaker~\cite{wang2023wespeaker} to extract the Speaker Embeddings of converted utterance and the target one, and compute the cosine Similarity between them which is denoted as SES.
We list the results of Ground Truth~(GT) as reference for the accuracy of FunASR and WeSpeaker.
Specifically, we select a few parallel sentences from ESD dataset for objective tests with ground truth. In Many-to-Many VC, ``SAVC" receives lower MCD values than baseline models. Even for unseen speakers, the proposed model still performs with less distortion. Considering that the recognition model has the problem of recognizing words with homophones and different synonyms, we also listed the recognition results of the Ground Truth. It's worth noting that the CERs is similar among PMVC, UUVC and SAVC in two VC tasks, which indicates that they all show a good intelligibility even with different prosody.



\subsection{Prosody Consistency Evaluation}
To further verify the expressiveness of each model, we select the baseline models with modeling of pitch-energy and conduct another test. Take the Pearson correlation coefficients of energy and pitch as prosody measurement metrics. A higher the Pearson Correlation Coefficient means the model is with more accurate prosody generation. 
The result of Tab.~\ref{table:PC} suggests that SAVC preserves the expressive qualities of the source speech well and makes similar performance to the model based on explicit prosody features.
\begin{table}[htbp]
   \centering
   \caption{Pearson correlation in f0 and energy.} 
    \label{table:PC}
    \begin{tabular}{l c c c c c}
     \hline
     Method & SAVC & UUVC &SpeechSplit2.0 &PMVC \\
     \hline
     $F0$ $\uparrow$ & \textbf{0.654} & 0.645 & 0.519 & 0.533\\
     $energy$ $\uparrow$ & \textbf{0.912} & 0.907 &  0.767 & 0.752\\
     \hline 
    \end{tabular}
\end{table}

\subsection{Ablation Study}
To investigate the performance of SAVC, we conduct three ablation experiments by dropping the adversarial style augmentation module~(ASA), prosody modeling~(pm), replacing soft speech units~(su) with discrete units respectively. Furthermore, a fake speech detection toolkit \textit{Resemblyzer}~\footnote{https: //github.com/resemble-ai/Resemblyzer} is used to conduct another objective experiment. Following VQ-CL~\cite{Tang2023VQ-CL}, the toolkit measure the similarity between ground truth and input utterances (5 real ones and 4 synthesized ones).
A detection score ranging from 0 to 1 is assigned to each converted on evaluation of voice similarity. The results are illustrated in Tab.~\ref{table:ablation}. When the ASA module is removed, the performance of the retrained model deteriorates in terms of voice similarity as the detection score decreases.
Dropping the prosody modeling by masking the prosody embedding results in noticeable degradation in overall voice quality. When the soft speech units are replaced with discrete units, the performance exhibits declines in naturalness and sound quality.


\begin{table}[htbp]
   \centering
   \caption{Objective and subjective results of different models } 
    \label{table:ablation}
    \begin{tabular}{l c c c}
     \hline
     Method & MCD $\downarrow$ & Detection Score$\uparrow$ &MOS $\uparrow$\\
     \hline
     SAVC & 5.45 $\pm$ 0.14 & 0.73 $\pm$ 0.08 & 3.73 $\pm$ 0.08\\
     w/o ASA & 5.53 $\pm$ 0.13 & 0.52 $\pm$ 0.07&  3.69 $\pm$ 0.17\\
     w/o su & 6.93 $\pm$ 0.09 & 0.72 $\pm$ 0.11&  3.57 $\pm$ 0.21\\
     w/o pm & 5.54 $\pm$ 0.05 & 0.68 $\pm$ 0.13&  3.72 $\pm$ 0.18\\
     \hline 
    \end{tabular}
\end{table}

\begin{figure}[!t]
    \centering
    \subfigure[Timbre simlarity]{
        \begin{minipage}[b]{0.9\linewidth}
        \label{Timbre simlarity}
        \centering
              \begin{tikzpicture}[scale = 0.5]
                \tikzstyle{every node}=[font=\small,scale=0.8]
                \pie[cloud,explode=0, text=inside, rotate = -20]
                {43.75/SAVC,9.38/M1,9.37/M2,31.25/M3,6.25/Fair } 
            \end{tikzpicture}
        \end{minipage}
    }
    \hfill
    \subfigure[Prosody simlarity]{
        \begin{minipage}[b]{0.9\linewidth}
        \label{Prosody simlarity}
        \centering
              \begin{tikzpicture}[scale = 0.5]
                \tikzstyle{every node}=[font=\small,scale=0.8]
                \pie[cloud, rotate = 0,radius=3.5]
                {37.50/SAVC, 31.25/M1, 21.88/M2, 6.25/M3,3.12/Fair} 
               
            \end{tikzpicture}
        \end{minipage}
    }
    
    \caption{Listening test results of ablation studies. (M1: w/o ASA; M2: w/o su; M3: w/o pm)}
    \label{abx}
\end{figure}
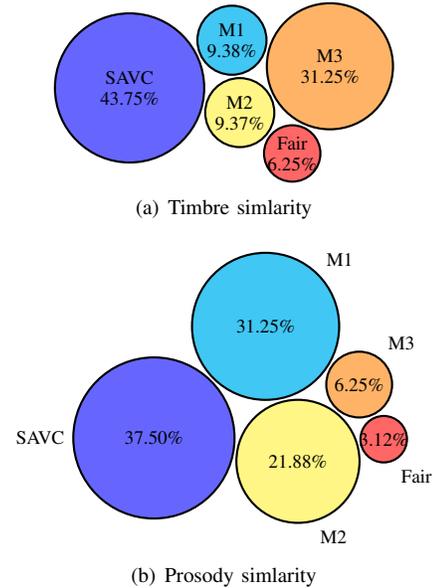

Besides, we ask 32 human volunteers to do blind listening tests to evaluate the timbre similarity and prosody similarity for ablation studies. In detail, given a real audio and fake ones from different models, the volunteers are asked which one is the most similar to the ground truth. If it's hard to decide, then choose the ``Fair" option.
As seen in Fig.~\ref{Timbre simlarity}, removing adversarial style augmentation degrades the speaker similarity. When the prosody modelling part is removed, the prosody similarity also decreases evidently as shown in Fig.~\ref{Prosody simlarity}.

\section{Conclusion}
In this paper, a novel framework is proposed to realize SSL feature-based expressive voice conversion task. Based on soft speech units we design a statistic perturbation to reshape speaker style. Then, the modeling of content and prosody by an attribute encoder independently. The experimental results show that the proposed method improves the sound quality of converted speech and the similarity to the target voice.
\section{Acknowledgement}
Supported by the Key Research and Development Program of Guangdong Province (grant No. 2021B0101400003). Corresponding author is Xulong Zhang (zhangxulong@ieee.org).
\clearpage

\bibliographystyle{IEEEtran}
\bibliography{reference}

\end{document}